\documentclass[5p,10pt]{elsarticle}

\usepackage{amsmath,amsthm,amscd,amssymb}
\def\alp{\alpha}
\def\be{\beta}
\def\ga{\gamma}

\def\si{\sigma}

\def\Ga{\Gamma}

\def\Si{\Sigma}
\def\Om{\Omega}

\usepackage[latin1]{inputenc}

\def\b1{{1\!\!1}}

\def\cR{{\ca R}}

\def\bC{{\mathbb C}}           

\def\bM{{\mathbb M}}

\def\gR{{\mathfrak R}}

\def\beq{\begin{eqnarray}}
\def\eeq{\end{eqnarray}}

\newcommand{\ca}[1]{{\cal #1}}         

\let\oo\o
\renewcommand{\o}{\perp}

\def\dnot{{\displaystyle{\not}}}
\newcommand{\mydef}{:=}

\newcommand{\fun}[1]{{\phantom{#1}}}

\def\vsp{\vspace{0.2cm}}

\newtheoremstyle{thm}
{12pt}
{12pt}
{\itshape}
{}
{\itshape\bfseries}
{}
{1em}
{}

\theoremstyle{thm}

\journal{Physics Letters B}

\begin{document}

\begin{frontmatter}


\title{A No-Go Theorem for the Consistent Quantization of Spin 3/2
  Fields on General Curved Spacetimes}



\author[label1]{Thomas-Paul Hack\corref{cor1}}
\ead{thomas-paul.hack@desy.de}
\author[label2]{Mathias Makedonski\corref{cor1}}
\ead{mathias.makedonski@math.ku.dk}
\cortext[cor1]{Corresponding author}
\address[label1]{II. Institut f\"ur Theoretische Physik, Universit\"at Hamburg,
Luruper Chaussee 149,
D-22761 Hamburg, Germany}
\address[label2]{Institut for Matematiske Fag, K\oo benhavns Universitet, Universitetsparken 5, 2100 Copenhagen, Denmark}


\begin{abstract}
It is well-known that coupling a spin $\frac32$-field to a
gravitational or electromagnetic background leads to potential
problems both in the classical and in the quantum theory. Various
solutions to these problems have been proposed so far, which are all
restricted to a limited class of backgrounds. On the other hand, negative results
for general gravitational backgrounds have been reported only for a
limited set of couplings to the background to date. Hence, to our
knowledge, a comprehensive analysis of all possible couplings to the
gravitational field and general gravitational backgrounds including
off-shell ones has not been performed so far. In this work we analyse
whether it is possible to couple a spin $\frac32$-field to a
gravitational field in such a way that the resulting quantum theory is
consistent on arbitrary gravitational backgrounds. We find that this
is impossible as all couplings require the background to be an
Einstein spacetime for consistency. This enforces the widespread belief that supergravity theories are the only meaningful models which contain spin $\frac32$ fields as in these models such restrictions of the gravitational background appear naturally as on-shell conditions.

\normalsize
\end{abstract}


\end{frontmatter}

\section{Introduction -- problems of spin $\frac32$ fields in non-trivial backgrounds}

A free spin $\frac32$-field $\psi$ of mass $m\ge0$ in flat four-dimensional Minkowski spacetime is described by the {\it Rarita-Schwinger equations} \cite{Rarita}

\begin{align}\label{raritaschwinger0}(\cR_0\psi)^\alp&\mydef(-i\dnot\partial+m)\psi^\alp
\\&\mydef(-i\ga^\mu\partial_\mu+m)\psi^\alp =0\notag\,,\\
\label{raritaschwinger00}\dnot\psi &\mydef \ga_\mu \psi^\mu=0\,.
 \end{align}
Here and in the following greek indices denote (co)\-tan\-gent space indices,
$\ga^\mu$ are the usual $\ga$-matrices, $\psi$ is a Dirac
spinor-valued vector field whose spinor indices we suppress throughout. Buchdahl realised already more than fifty years ago that a
minimal coupling of the above equation to a background gravitational field leads to problems \cite{Buchdahl0}: the minimally coupled equations imply $R_{\mu\nu}\ga^\mu\psi^\nu=0$, with $R_{\mu\nu}$ denoting the Ricci curvature tensor, and this equation can only be satisfied by $\psi\equiv 0$ unless the spacetime is an Einstein spacetime s.t. $R_{\mu\nu}$ is a constant multiple of the metric $g_{\mu\nu}$.

Later Johnson and Sudharsan found that the quantum theory of a spin
$\frac32$-field minimally coupled to an electromagnetic background
field fails to satisfy unitarity \cite{Johnson}. This result has been
complemented by Velo and Zwanziger who pointed out that the coupling to an electromagnetic
field is already problematic at the classical level as it leads to superluminal propagation \cite{Velo}. 

This last finding seemed to be the most shocking as it became famous as the {\it Velo-Zwanziger problem}.

All three problems have been analysed in great detail and various
solutions have been proposed. As it is impossible to provide a
comprehensive list of earlier works, we only mention a few selected
ones. Special, i.e. maximally symmetric or constant gravitational and
electromagnetic backgrounds have been studied e.g. in \cite{Deser2,
  Deser2a, Deser2b, Porrati} where the causality and/or unitarity problems have been proven to be absent for special values of the mass
and/or the couplings. In \cite{Madore, Deser3} it was pointed out that
all problems can be solved in Einstein-Maxwell backgrounds at the cost
of very small or very large masses $m$. The most prominent solution of
the Buchdahl-problem is arguably supergravity \cite{Deser, Freedman},
where the Einstein condition on the spacetime appears as a natural
on-shell condition. The causal behaviour of supergravity was shown in
\cite{Choquet}, whereas unitarity had mostly been discussed on
maximally symmetric Einstein backgrounds such as Minkowski and Anti de
Sitter spacetime \cite{Deser2, Deser2a, Deser2b}. Recently unitarity
has also been proven for general, asymptotically flat and Ricci flat
Einstein backgrounds\footnote{In a previous preprint version of this
  work we had argued that supergravity fails to satisfy unitarity on
  the basis of a spin $\frac32$ field equation derived from the
  original equation of motion in supergravity. In \cite{Schenkel} it
  was pointed out that our argument fails if one considers the original
  supergravity equations of motion instead.} \cite{HS}. Other solutions to the Buchdahl
problem, which avoid restrictions on the background, have been
proposed and analysed both in the $(1,\frac12)\oplus(\frac12,1)$ representation, e.g. \cite{Frauendiener, Baer2, Muehlhoff3}, and in the
$(\frac32,0)\oplus(\frac12,1)$ representation of $SL(2,\bC)$ \cite{Buchdahl2, Buchdahl3, Wuensch, 
 Muehlhoff1, Mathias}. While the
former suffer either from the causality or the unitarity problem, the
latter satisfy causality, but a unitarity proof is lacking to date.

All the above-mentioned analyses have in common that they consider
restrictions on the couplings, the mass, or the background
fields. Whereas in \cite{Deser4} general non-minimal couplings to the
electromagnetic field have been studied with a negative result, it
seems that a comprehensive study of general non-minimal couplings to
the gravitational field and general gravitational backgrounds has not
been available to date. In this letter, we thus investigate whether it
is possible to couple a spin $\frac32$-field to a gravitational field
in a way, such that the resulting quantum theory is causal, unitary,
and propagates the correct degrees of freedom on arbitrary spacetime
backgrounds -- including off-shell ones. This generality is motivated by
the modern approach to quantum field theory on curved spacetimes
\cite{BFV} (see also \cite{Thomas} for an extensive review) where one
tries to quantize a model without using any knowledge on the
background spacetime other than its defining properties such as
e.g. the Lorentzian metric signature. As this turns out to be possible
for spins $\le 1$, see e.g. \cite{HW01, BFV, Hollands07, Sanders3,
  DHP, DalaiLama, Fredenhagen}, the question, whether this is the case for higher spins as well,
naturally arises. However, we find that a
background-independent consistent quantization seems to be impossible
for spin $\frac32$-fields in gravitational backgrounds.

We work solely in the $(1,\frac12)\oplus(\frac12,1)$
representation of $SL(2,\bC)$ and do not consider the
$(\frac32,0)\oplus(\frac12,1)$ representation, which is equivalent to
the former on flat spacetimes, but not on curved ones. This is
motivated by the results of \cite{Mathias} where it has been found
that unitarity of a quantum field in this representation 
is unlikely to hold due to its very structure in curved spacetimes.

Our letter is organised as follows. In section \ref{sconditions} we
compile four conditions which a consistent quantum theory of a spin
$\frac32$-field on an arbitrary curved spacetime should satisfy. While
the causality condition and the condition on the degrees of freedom
are well-known, the very ``background independence'' condition has
apparently not been discussed so far in this context. Our fourth
condition, a certain symmetry condition of the field equations, is
shown to be virtually equivalent to unitarity and thus replaces the unitarity
condition. Furthermore we point out that, in contrast to
statements in the literature, causality and unitarity are not
equivalent for spin $\frac32$ fields. In section \ref{nogo} we finally prove our no-go theorem and show that no non-minimally coupled spin $\frac32$-field equation satisfies all four conditions. The letter ends with a discussion of our findings in section \ref{discussion}.

\section{Conditions for a consistent spin $\frac32$-quantum theory in curved spacetimes}
\label{sconditions}

We consider a spin $\frac32$-field $\psi$ on a general curved
spacetime $(M,g_{\mu\nu})$, i.e. $M$ is a four-dimensional manifold,
$g_{\mu\nu}$ a metric with signature $(+,-,-,-)$ and $\psi$ is a
four-spinor-valued vector field whose vector index we shall write only if necessary. We shall often denote $(M,g_{\mu\nu})$ by $M$ for simplicity. The field equations for $\psi$ are 
\begin{gather}\label{raritaschwinger}\cR\psi=0\,,\\
\label{raritaschwinger1}\dnot\psi\mydef \ga_\mu \psi^\mu=A_\mu \psi^\mu\,,\end{gather}
where $\cR$ is an arbitrary first order differential operator constructed out of the metric, the curved-spacetime $\ga$-matrices $\ga^\mu$, and the mass $m$, and $A_\mu$ is an arbitrary zeroth order operator of that kind. Thus, with tuples $(\cR, A_\mu)$ we parametrise all non-minimal couplings of $\psi$ to the background gravitation field. By $S(\cR,M)$ we denote the set of all (infinitely often differentiable) solutions of \eqref{raritaschwinger} on the spacetime $M$, whereas by $S(\cR,A_\mu,M)$ we denote the subset of $S(\cR,M)$ which satisfies in addition \eqref{raritaschwinger1}. We now list four conditions on $(\cR, A_\mu)$ and argue why they sufficient for a spin $\frac32$-quantum theory induced by $(\cR, A_\mu)$ to be consistent in arbitrary curved spacetimes.

\subsection{Condition 1: Irreducibility}
\vsp
{\it \noindent On Minkowski spacetime $\bM$, $A_\mu\equiv 0$ and $S(\cR,0,\bM)=S(\cR_0,0,M)$.}\\\\
This condition requires that $(\cR, A_\mu)$ define a theory which
propagates the correct number of degrees of freedom for a spin
$\frac32$-field of mass $m$. This is here achieved by comparison with
the standard theory in Minkowski spacetime, which after all is the
very spacetime in which the concepts of ``spin'' and ``mass'' are
defined via irreducible representations of the Poincar\'e group. We
don't require $\cR\equiv \cR_0$ on $\bM$ because different $\cR$ can
be equivalent on-shell.

\subsection{Condition 2: Causality}
\vsp
{\it \noindent $\cR$ is hyperbolic and the constraint $\dnot\psi=A_\mu \psi^\mu$ is compatible with time evolution.}\\\\
Hyperbolic field equations such as the Klein-Gordon or the Dirac
equation guarantee causal propagation of the degrees of freedom, see
e.g. \cite{DreiFrauen, Courant, Madore, Baer}, as they limit the
dependence of a solution $\psi(x)$ at a point $x$ to the past
lightcone of $x$. Hyperbolicity is a condition on the coefficient
matrix $\si^\mu$ of the highest derivative term $\si^\mu \nabla_\mu$
in $\cR$, the so-called {\it principal symbol}: for a spacelike/timelike vector $k_\mu$, $k_\mu \si^\mu$ must be invertible, while for a lightlike $k_\mu$, it must have vanishing determinant. Additionally, the above compatibility condition is required to avoid that $S(\cR,A_\mu,M)$ contains only the trivial solution $\psi\equiv 0$.

\subsection{Condition 3: Background independence}
\vsp
{\it \noindent The number of degrees of freedom propagated by
  $(\cR,A_\mu)$ is independent of the background spacetime
  $M$. Moreover, either $A_\mu\equiv 0$ on all spacetimes, or
  \eqref{raritaschwinger1} is automatically satisfied for all
  solutions of \eqref{raritaschwinger}.}\\\\
This condition is required to avoid the Buchdahl-problem mentioned in section 1, where it happens that the minimally coupled Rarita-Schwinger equations \eqref{raritaschwinger0} and \eqref{raritaschwinger00} propagate the correct number of degrees of freedom on Einstein spacetimes, but no degrees of freedom at all otherwise.

Stated in more technical terms this condition requires
that $S(\cR,A_\mu,M)$ is {\it locally contravariant} in the sense of
\cite{BFV}: if we consider two spacetimes $M_1\subset M_2$ where one
is a (suitable) subset of the other, then $S(\cR,A_\mu,M_1)$ should be
equal to the restriction of $S(\cR,A_\mu,M_2)$ to $M_1$.

We impose the additional condition on $A_\mu$ because we have not been
able to prove that the constraint \eqref{raritaschwinger1} satisfies
our background-independence condition except in these two special cases. 

\subsection{Condition 4: Selfadjointness}
Although this condition appears to be the most technical one, it is
equivalent to demanding that the field equation
\eqref{raritaschwinger} can be obtained from a quadratic action. We state the condition first and comment on its relation to unitarity afterwards. To this avail, we introduce the notion $\Ga_0(M)$ for the set of (infinitely often differentiable) vector-spinor valued functions which vanish outside of a compact subset of $M$, so-called {\it test functions}. For two test functions $f_1$, $f_2$, we define a product $\langle f_1, f_2\rangle$ by
$$\langle f_1, f_2\rangle \mydef \int\limits_M d^4x\;\sqrt{-\det g_{\mu\nu}}\; g_{\alp\be}\overline{f^\alp_1(x)}f^\be_2(x)\;,$$
where the bar denotes the usual Dirac conjugation of a four-spinor. We can define the {\it adjoint} $\cR^\dagger$ of $\cR$ with respect to $\langle\cdot,\cdot\rangle$ by $\langle \cR^\dagger f_1, f_2\rangle\mydef \langle  f_1, \cR f_2\rangle$ and finally state the fourth and last condition.\\\\
{\it \noindent $\cR$ is formally selfadjoint: $\cR^\dagger = \cR$, i.e. $\langle \cR f_1, f_2\rangle=\langle  f_1, \cR f_2\rangle$.}\\\\
To discuss the relation of this condition to unitarity, we briefly recall the unitarity condition for a spin $\frac32$-field, see e.g. \cite{Johnson, Deser2b, HS} for details. To wit, the {\it covariant anticommutator} of the quantized field $\psi$ and its adjoint $\overline\psi$ is after canonical quantization given by
\begin{equation}\label{car}\{\psi(x),\overline\psi(y)\}=i G(x,y)\end{equation}
where $G(x,y)$ is the so-called {\it anticommutator function}, a
generalisation of the {\it Pauli-Jordan-function} for scalar
fields. $G(x,y)$ is equal to the difference of the advanced and
retarded Green's function\footnote{For a hyperbolic $\cR$, these
  Green's functions exist and are unique on any spacetime which fulfils the
  so-called {\it
  global hyperbolicity} condition, see \cite{Baer, Baer2} for details;
this quite natural condition on $M$ shall be tacitly assumed throughout this letter.} of the differential operator $\cR$ and thus
$G(x,y)$ depends on the specific form of $\cR$ and satisfies $\cR_x
G(x,y) = \cR^\dagger_y G(x,y)=0$ for a general hyperbolic $\cR$. The operator $G$ defined by
$$[G f](x)\mydef \int\limits_M d^4y\;\sqrt{-\det g_{\mu\nu}}\; G(x,y)f(y)\;,$$
maps test functions to solutions which have finite spatial extent at each time, i.e. ``wave packets''. Accordingly, the quantized field $\psi(x)$ integrated with the Dirac adjoint of a test section $f$ -- henceforth denoted by $\psi(\overline f)$ -- can be interpreted as the quantum operator corresponding to the classical wave packet $Gf$. Physical wave packets should satisfy the constraint $\ga_\mu (Gf)^\mu = A_\mu(Gf)^\mu$ in addition to the equation $\cR (Gf)=0$ and we denote the corresponding ``physical subspace'' of the test sections $\Ga_0(M)$ by $\Ga_0(\cR, A_\mu, M)$. If one now considers the anticommutation relations \eqref{car} integrated with a test section $f\in \Ga_0(\cR, A_\mu, M)$ and its Dirac adjoint
$$\label{car}\{\psi(\overline f),\overline\psi(f)\}=i G(\overline f,f)=i\langle f, Gf \rangle\,,$$ 
then the right hand side must be a positive number because the left
hand side is of the form $B^\dagger B + B B^\dagger$ with
$B=\overline\psi(f)$ and thus has positive expectation value in any
quantum state $|\Om\rangle$. Hence, the non-trivial unitarity
condition for the tuple $(\cR, A_\mu)$ is that the anticommutator
function $G(x,y)$ determined by $\cR$ must satisfy $$i\langle f,
Gf\rangle\ge0$$ for any physical test function $f\in \Ga_0(\cR, A_\mu,
M)$. Note that, for a formally selfadjoint $\cR$ the 
previously discussed {\it covariant anticommutation} relations are equivalent to {\it equal-time anticommutation relations}, see e.g. \cite{Thomas, Baer2, HS} for details. Basically this follows from the identity
\begin{equation}\label{equaltime} \langle f_1, Gf_2\rangle =\int\limits_\Si\! d^3x  \sqrt{-\det h_{ij}} \overline{Gf_1} n_\mu \si^\mu Gf_2\,.\end{equation}
where $\Si$ is an arbitrary equal-time surface of $M$ with normal vector $n_\mu$ and $h_{ij}$ is the spatial metric on $\Si$ induced by $g_{\mu\nu}$.

We shall now demonstrate the close relation between the
selfadjointness condition $\cR^\dagger = \cR$ and the unitarity
condition $i\langle f, Gf\rangle\ge0$ which lead us to replace the
latter, which is difficult to check directly on all spacetimes, with the former, which can be checked more easily.

To start with, we shall argue why the selfadjointness condition
implies unitarity on any topologically trivial spacetime $M$ if
unitarity is known in Minkowski spacetime $\bM$. To see this, we
consider any topologically trivial spacetime $M$ and deform it in such
a way that it becomes Minkowski in the past, see \cite{Fulling} for
details. Loosely speaking, we consider a fiducial spacetime
$M^\prime$ such that the metric
on $M^\prime$ equals the metric on $M$ for large
positive times, whereas for large negative times it equals the
Minkowski metric. Given such a deformation and a formally selfadjoint
$\cR$, the identity \eqref{equaltime} allows us to compute $\langle f,
G f\rangle$ on any equal-time surface of $M^\prime$, in particular
also in the Minkowski region where we know that it is positive by
assumption. Moreover, for the equations \eqref{raritaschwinger0} and \eqref{raritaschwinger00}, unitarity can be easily checked by an explicit computation in Fourier space, thus our first condition together with selfadjointness is sufficient to guarantee unitarity on any topologically trivial $M$.

We now prove that $i\langle f, Gf\rangle\ge0$ for $f\in \Ga_0(\cR,A_\mu,M)$ implies
$\langle f_1, \cR f_2\rangle=\langle \cR f_1, f_2\rangle$ for $f_i\in
\Ga_0(\cR,A_\mu,M)$ on arbitrary spacetimes. Defining a product on
physical test functions by $(f_1,f_2)\mydef i\langle f_1, Gf_2\rangle$, our
assumption $(f,f)\ge 0$ implies by polarisation that the complex
conjugate of $(f_1,f_2)$ equals $(f_2,f_1)$ from which we can deduce
that $iG$ is formally selfadjoint on $\Ga_0(\cR,A_\mu,M)$. As
$G^\dagger$ is the operator corresponding to the anticommutator
function of $\cR^\dagger$, we find that $G^\dagger = G$ on physical
test functions and the same is true for the advanced $G^{(\dagger)}_+$ and
retarded $G^{(\dagger)}_-$ pieces of $G$ and $G^\dagger$
respectively because these are unique. Using this, $\cR G_\pm = G_\pm \cR = 1$ and the fact that
$\cR$ maps $\Ga_0(\cR,A_\mu,M)$ to itself we can compute $$\cR^\dagger
f = \cR^\dagger G_\pm \cR f = \cR^\dagger G^\dagger_\pm \cR f = \cR
f\,.$$

In order for the general selfadjointness condition to be equivalent to the
unitarity condition for the purposes of a no-go theorem, it would be
necessary to prove that unitarity implies selfadjointness of $\cR$ on
all test functions and not only on the physical ones. Alternatively,
we could also require the latter, weaker selfadjointness
condition. However, one could just as well argue that the stronger, 
general selfadjointness condition is important in its own right
irrespective of unitarity because it is equivalent to demand that
$\cR$ comes from a quadratic action. Thus, we proceed with this
stronger condition, because it is easier to verify.

\section{A no-go theorem for the consistent quantization of
  non-minimally coupled spin $\frac32$-fields on general curved spacetimes}
\label{nogo}

We shall prove in the following that a large class of non-minimally coupled field
equations $(\cR, A_\mu)$ does not satisfy the four conditions compiled
in the previous section. In the course of proving this no-go theorem,
it will become clear that the proof can be extended to any larger
class of operators without much effort, such that the class we shall
consider can be safely regarded as effectively exhausting all possible
covariant field equations in the $(1,\frac12)\oplus (\frac12, 1)$
representation of $SL(2,\bC)$.

To wit, we consider $\cR$ of the form
\begin{align*}
(\cR\psi)^\alp&\mydef\left(-i\dnot\nabla+m\right)\psi^\alp+a_0
m\ga^\alp \dnot \psi+a_1 i\nabla^\alp\dnot \psi  \\&\quad +a_2 i\ga^\alp
\nabla_\mu\psi^\mu+a_3 i\ga^\alp \dnot \nabla \dnot \psi
+\widetilde\psi^\alp\\
\widetilde\psi^\alp&\mydef m \ga^\alp B+m C^\alp+iD^\alp+i\ga^\alp E\\
B&\mydef b_1 R_{\mu\nu}\ga^\mu\psi^\nu+b_2 R\dnot\psi\\
C^\alp&\mydef c_1R^\alp_{\fun{\alp}\nu}\psi^\nu + c_2
R^\alp_{\fun{\alp}\nu} \ga^\nu \dnot \psi + c_3 R\psi^\alp\\&\quad+c_4
\gR^\alp_{\fun{\alp}\nu} \psi^\nu
\end{align*}
\begin{align*}
D^\alp&\mydef d_1 R^\alp_{\fun{\alp}\nu} \dnot \psi^\nu + d_2 \left(\dnot\nabla R^\alp_{\fun{\alp}\nu}\right)\psi^\nu+d_3 R^\alp_{\fun{\alp}\nu} \ga^\nu \dnot\nabla\dnot\psi\\&\quad + d_4 \left(\dnot \nabla R^\alp_{\fun{\alp}\nu}\right)\ga^\nu \dnot \psi + d_5 R\dnot\nabla \psi^\alp + d_6 \left(\dnot \nabla R\right)\psi^\alp\\
 &\quad+ d_7 R^\alp_{\fun{\alp}\nu}\nabla^\nu \dnot\psi + d_8 \left(\nabla^\alp R\right)\dnot \psi + d_9 R\nabla^\alp \dnot \psi\\&\quad + d_{10}\gR^\alp_{\fun{\alp}\nu} \nabla^\nu \dnot \psi + d_{11}\left(\nabla^\nu \gR^\alp_{\fun{\alp}\nu}\right)\dnot \psi \\&\quad+ d_{12} R_{\mu\nu} \nabla^\alp \ga^\mu \psi^\nu
 + d_{13}\left(\nabla^\alp R_{\mu\nu}\right)\ga^\mu\psi^\nu  \\&\quad+ d_{14}\left(\nabla_\nu R^\alp_{\fun{\alp}\mu}\right)\ga^\mu \psi^\nu + d_{15}R^\alp_{\fun{\alp}\mu} \ga^\mu \nabla_\nu \psi^\nu \\
 E&\mydef e_1 R_{\mu\nu}\ga^\nu \dnot\nabla \psi^\nu + e_2 \left(\dnot \nabla R_{\mu\nu}\right)\ga^\mu\psi^\nu + e_3 R\dnot\nabla \dnot\psi\\&\quad + e_4\left(\dnot\nabla R\right)\dnot\psi+ e_5 \left(\nabla_\nu R\right)\psi^\nu + e_6 R \nabla_\nu \psi^\nu\\
 &\quad + e_7 \left(\nabla^\mu \gR_{\mu\nu}\right)\psi^\nu + e_8 \gR_{\mu\nu}\nabla^\mu\psi^\nu + e_9 R_{\mu\nu}\nabla^\mu\psi^\nu\,.
\end{align*}
where $\nabla_\mu$ is the spin covariant derivative, $a_i \in \bC$ are arbitrary constants whereas $\gR_{\alp\be}=\frac14 R_{\alp\be\mu\nu}\ga^\mu\ga^\nu$ denotes
the spin curvature tensor\footnote{Note that all couplings containing
  the Riemann tensor $R_{\alp\be\mu\nu}$ can be expressed via the spin
  curvature tensor $\gR_{\alp\be}$. Furthermore, we have omitted all
  couplings which would be linearly dependent by means of Bianchi
  identities. We follow \cite{Wald} regarding conventions in the
  definition of the curvature tensors.}. Moreover, derivatives in
parenthesis are meant to act only on the jointly enclosed curvature tensors, and $b_i$, $c_i$, $d_i$, $e_i$ are arbitrary complex-valued functions of curvature invariants and $m$ of mass dimension $-2$.

We start our proof by checking selfadjointness, since this turns out
to be the strongest condition. Indeed, as one can check by direct
computation, it is fulfilled on arbitrary curved
spacetimes if and only if the following equations are true.
$$a^*_0=a_0\qquad a_2 = a^*_1 \qquad a^*_3 = a_3\qquad b_1 = c^*_2$$$$ b^*_2=b_2\qquad c^*_1=c_1\qquad c^*_3=c_3\qquad c^*_4=c_4$$
$$d_1=d_3=d_5=d_7=d_9=d_{10}=d_{12}=d_{15}=0$$$$e_1=e_3=e_6=e_8=e_9=0$$
$$d^*_2=d_2\qquad d^*_4=e_2\qquad d^*_6=d_6\qquad d_8 = e^*_5$$$$ d_{11}=e^*_7 \qquad d^*_{13}=d_{14}\qquad e^*_4=e_4$$
Here, $^*$ denotes complex conjugation. In essence, requiring $\cR^\dagger=\cR$ rules out terms where a curvature tensor multiplies a derivative of $\psi^\alp$, because such terms generate derivatives of curvature tensors by the partial integration involved in the definition of the formal adjoint of $\cR^\dagger$. These curvature tensor derivatives can not be cured by explicitly adding couplings of $\psi^\alp$ to curvature derivatives, as such terms must be present both in $\cR$ and in $\cR^\dagger$. Hence, selfadjointness rules out {\it arbitrary} terms where a curvature tensor multiplies a derivative of $\psi^\alp$, extending the validity of this proof to a larger class of $\cR$ containing all possible such terms. 

We proceed by checking the hyperbolicity bit of of our causality
condition. Let $k_\mu$ be timelike or spacelike and let $\psi^\alp$ fulfil
$$ik_\mu\si^\mu \psi^\alp = \dnot k \psi^\alp - a_1 k^\alp \dnot \psi - a_2 \ga^\alp k_\mu \psi^\mu - a_3 \ga^\alp \dnot k \dnot \psi=0 \,,$$
where we have already taken into account that the allowed principal symbols are reduced by selfadjointness. We have to check for which $a_i$ the above equation implies $\psi^\alp\equiv 0$. By multiplying the above equation with $\dnot k$ and $k^\alp$, we can obtain the following derived equations
$$(1-a_2)\dnot k k_\mu\psi^\mu = (a_1+a_3)k^2\dnot \psi$$$$ (1-3a_2)\dnot k k_\mu\psi^\mu = (1+3 a_3)k^2\dnot \psi\,,$$
which can be rewritten as 
$$\left(\begin{array}{cc}(1-a_2)\b1 & -(a_1+a_3)\b1\\(1-3a_2)\b1&
    -(1+3 a_3)\b1\end{array}\right)\left(\begin{array}{c}\dnot k
    k_\mu\psi^\mu\\k^2\dnot \psi\end{array}\right)=0\,,$$
where $\b1$ is the $4\times4$ identity matrix.
As $k_\mu$ is timelike or spacelike, this equation together with
$ik_\mu\si^\mu \psi^\alp=0$ implies $\psi^\alp\equiv 0$ if and only if the determinant of the appearing $8\times 8$ matrix is non-zero; this in turn is the case iff
\beq\label{detnonzero}-3 a_1 a_2 + a_1+a_2-2a_3-1 \neq 0\,.\eeq
We do not discuss lightlike $k_\mu$, as \eqref{detnonzero} will be sufficient to prove the theorem.

Finally, we verify the background-independence and irreducibility
conditions. To this avail, we contract $(\cR\psi)^\alp = 0$ with both
$\ga_\alp$ and $\nabla_\alp$ and combine the results to obtain the following equation for $\dnot \psi$.
\begin{align}
-\left(\frac{(a_2-1)(1+a_2+4a_3)}{2-4 a_2}+a_1+a_3\right)\nabla_\mu\nabla^\mu \dnot
\psi\notag\\ 
+ \left(\frac{(a_2-1)(1+4a_0)}{2-4 a_2}+\frac{1+a_1+4a_3}{2-4a_2}+a_0\right)im\dnot\nabla \dnot\psi \notag\\
+\left(\frac{(a_2-1)(1+a_2+4a_3)}{2-4 a_2}+a_3\right)\frac R4
\dnot\psi \notag\\
+\frac{1+4 a_0}{2-4 a_2}m^2 \dnot\psi - \frac12 R_{\mu\nu}\ga^\mu\psi^\nu\notag\\
+\frac{a_2-1}{2-4 a_2}i\dnot \nabla \dnot\widetilde\psi + i\nabla_\mu \widetilde \psi^\mu + \frac{m}{2-4 a_2}\dnot\widetilde\psi=0\,.\label{eqnot}
\end{align}
Here, our first condition assures that $2-4 a_2\neq 0$. To see this,
note that contracting $\cR\psi^\alp = 0$ with $\ga_\alp$ yields an
equation which can be rewritten as 
\begin{gather}(2-4 a_2)i\nabla_\mu \psi^\mu\notag\\=(1+a_1+4a_3)i\dnot\nabla\dnot \psi + (1+4a_0)m\dnot\psi + \dnot\widetilde\psi\,. \label{divergence}\end{gather}
If $2-4 a_2=0$, then $\nabla_\mu \psi^\mu=0$ would not follow from
$\cR\psi^\alp = 0$ and $\dnot \psi=0$ on Minkowski spacetime, hence
$\dnot S(\cR,0,\bM)= \dnot S(\cR_0,0,\bM)$ would not hold because all
elements of $\dnot S(\cR_0,0,\bM)$ satisfy $\nabla_\mu \psi^\mu=0$.

To assure that our background independence condition holds, we have to
either guarantee that $\dnot\psi^\alp = A_\mu\psi^\mu$ holds
automatically for solutions of $\cR\psi = 0$ or that $A_\alp\equiv 0$
on all spacetimes. Let us check if the first of these conditions can
be fulfilled. Without specifying $A_\mu$ explicitly, we know that, in
Minkowski spacetime, $A_\mu\equiv 0$ must hold on account of the
irreducibility condition. However, in flat spacetime, \eqref{eqnot} is
a hyperbolic partial differential equation for $\dnot\psi$, as the
coefficient of $\nabla_\mu\nabla^\mu \dnot \psi$ is non-zero if we
apply the condition \eqref{detnonzero} derived from causality and
selfadjointness. Such a differential equation has certainly more
possible solutions than just $\dnot\psi\equiv0$, hence, by combining
causality, selfadjointness, and irreducibility, we find that only the
optional background independence condition that $A_\mu$ be identically vanishing on all spacetimes can be fulfilled. Inserting this into \eqref{eqnot}, we are left with
\begin{gather}\notag-\frac12 R_{\mu\nu}\ga^\mu\psi^\nu-\frac{a_2-1}{2-4 a_2}i\dnot \nabla \dnot\widetilde\psi \\+ i\nabla_\mu \widetilde \psi^\mu + \frac{m}{2-4 a_2}\dnot\widetilde\psi=0\,. \label{eqnot2}\end{gather}
In Minkowski spacetime, this equation is identically fulfilled and,
hence, poses no additional constraints on solutions of $\cR \psi=0$
and $\dnot\psi = 0$. To check if our background independence holds, we have to make sure that \eqref{eqnot2} is identically fulfilled on {\it all} spacetimes once $\cR \psi^\alp=0$ and $\dnot\psi = 0$ hold. To this avail, we insert $\dnot \psi = 0$ into \eqref{divergence}, and both $\dnot \psi=0$ and \eqref{divergence} into $\cR\psi^\alp = 0$ to obtain
$$i\nabla_\mu \psi^\mu = \frac{1}{2-4a_2}\dnot\widetilde\psi\,,$$$$(-i\dnot \nabla + m)\psi^\alp + \frac{a_2}{2-4 a_2}\ga^\alp \dnot\widetilde\psi + \widetilde\psi^\alp = 0\,.$$
These two equations are the only information on first derivatives of
$\psi^\alp$ 
one can obtain from $\cR \psi=0$ and $\dnot\psi = 0$. However, the
summand $\nabla_\mu \widetilde \psi^\mu$ in \eqref{eqnot2} contains
first derivatives of $\psi^\alp$ also in terms like e.g. $R_{\mu\nu}
\nabla^\mu\psi^\nu$, on which $\cR \psi=0$ and $\dnot\psi = 0$ give no
information in general curved spacetimes. Hence, these terms must
identically vanish in $\nabla_\mu \widetilde \psi^\mu$, which implies
that the coefficients of all terms in $\widetilde\psi^\alp$ surviving the insertion of $\dnot\psi = 0$ and whose free index $^\alp$ does not belong to $\ga^\alp$ or $\psi^\alp$ must vanish. Moreover the coefficients of all terms where $\ga^\alp$ appears followed by other $\ga$-matrices must vanish as well, as these terms also give rise to terms like e.g. $R_{\mu\nu} \nabla^\mu\psi^\nu$ if one considers them in $\nabla_\mu \widetilde \psi^\mu$ and commutes the contracted covariant derivative $\dnot\nabla$ with the additional $\ga$-matrices in order to use the available information on $\dnot\nabla \psi^\alp$. Analogously, the terms in $\widetilde\psi^\alp$ where the free index $^\alp$ belongs to $\psi^\alp$ but $\psi^\alp$ is multiplied by $\ga$-matrices are problematic in $\dnot \nabla \dnot\widetilde\psi$ and have to vanish identically. Altogether, avoiding the appearance of in general undetermined $\psi^\alp$-derivatives in \eqref{eqnot2} enforces
$$b_1=c_1=c_4=d_2=d_6=d_{13}=d_{14}=e_2=e_7=0\,,$$
hence, the remaining terms in $\widetilde \psi^\alp$ not yet ruled out
by background-independence are
$$\widetilde\psi^\alp = mc_3 R\psi^\alp + e_5 \ga^\alp \left(\nabla_\nu R\right)\psi^\nu\,.$$
We can now explicitly compute the left hand side of \eqref{eqnot2} by
inserting this expression for $\widetilde\psi^\alp$ and the knowledge
on $\nabla_\mu \psi^\mu$ and $\dnot\nabla \psi^\alp$ obtained from
$\cR\psi = 0$ and $\dnot\psi=0$. The result does not contain any
derivatives of $\psi^\alp$, but is a sum of various curvature tensors multiplying $\psi^\alp$. In general spacetimes, some of these terms are linearly independent and, hence, have to vanish individually in order for \eqref{eqnot2} to be identically fulfilled on all spacetimes. Particularly, since the only term in the left hand side of \eqref{eqnot2} containing the Ricci tensor turns out to be the one explicitly visible in \eqref{eqnot2}, we obtain
$$R_{\mu\nu}\ga^\mu\psi^\nu=0$$
as a necessary condition for \eqref{eqnot2} to hold on general
spacetimes. However, this is in conflict with background-independence, which closes the proof.

One can imagine that the steps taken in the last paragraph of this
proof can be generalised to arbitrary couplings of the curvature to
$\psi^\alp$, and we have argued in the discussion of selfadjointness
that the same holds for arbitrary couplings of the curvature to
derivatives of $\psi^\alp$, hence, we presume that our proof
effectively exhausts {\it all} possible covariant first order
differential operators $\cR$. Finally, we would like to emphasise that our proof covers both $m>0$ and $m=0$.

\section{Discussion}
\label{discussion}
The proof of our no-go theorem shows that, even if one allows for a
spin $\frac32$-field in a gravitational background to be coupled to
the gravitational field in an arbitrary non-minimal way, one is lead
to the same Buchdahl-problem present for the minimally coupled
equations of motion if one requires in addition that causality and
unitarity hold: the model is, at best, only consistent on Einstein
spacetimes. Whereas this seems to be a very restrictive condition for
the consistent quantization of spin $\frac32$-fields on curved
backgrounds, it fits nicely into the widespread picture that
supergravity theories are the only consistent models which contain
elementary spin $\frac32$-fields, see e.g. \cite{Esole}, as in these
models such conditions on the background appear naturally as on-shell
conditions \cite{Deser, Freedman}. One can expect that a
generalisation of our no-go theorem to the case where scalar and vector
background fields are present in addition to the metric field yields
conditions on the background which are compatible with on-shell
conditions in extended supergravity models, see e.g. \cite{Kofman3}.

\vspace*{9mm}
{\noindent\bf Acknowledgements}\\
T.P.H. gratefully acknowledges financial support from the Hamburg research cluster LEXI ``Connecting Particles with the Cosmos''. It is a pleasure to thank Wilfried Buchm\"uller, Claudio Dappiaggi, Stanley Deser, Klaus Fredenhagen, Jan Heisig, Stefan Hollands, Katarzyna Rejzner, Alexander Schenkel, Daniel Siemssen, Christoph Uhlemann and Jochen Zahn for valuable comments and illuminating discussions.

\end{document}